\documentclass[prb,aps,twocolumn,showpacs]{revtex4}
\usepackage{graphicx}
\usepackage{amsmath,amssymb,epsf,bm}
\begin{document}

\title{Spin Current Generation and Detection in the Presence of AC Gate}
\author{A. G. Mal'shukov$^{1}$, C. S. Tang$^{2}$, C. S. Chu$^{3}$ and K. A.
Chao$^{4}$}
\affiliation{$^{1}$Institute of Spectroscopy, Russian Academy of Science,
142190, Troitsk, Moscow oblast, Russia \\
$^2$Physics Division, National Center for Theoretical Sciences, P.O. Box 2-131,
Hsinchu 30013, Taiwan \\
$^{3}$Department of Electrophysics, National Chiao-Tung University, Hsinchu
30010, Taiwan \\
$^{4}$Solid State Theory Division, Department of Physics, Lund University,
S-22362 Lund, Sweden}

\begin{abstract}
We predict that in a narrow gap III-V semiconductor quantum well or quantum
wire, an observable electron spin current can be generated with a time
dependent gate to modify the Rashba spin-orbit coupling constant. Methods to
rectify the so generated AC current are discussed. An all-electric method of
spin current detection is suggested, which measures the voltage on the gate
in the vicinity of a 2D electron gas carrying a time dependent spin current.
Both the generation and detection do not involve any optical or magnetic
mediator.
\end{abstract}

\pacs{71.70.Ej, 72.25.Dc, 73.63.-b}
\maketitle

One key issue in spintronics based on semiconductor is the
efficient control of the spin degrees of freedom. Datta and
Das~\cite{datta} suggested the use of gate voltage to control the
strength of Rashba spin-orbit interaction (SOI)~\cite{rashba}
which is strong in narrow gap semiconductor heterostructures. In
InAs-based quantum wells a variation of 50\% of the SOI coupling
constant was observed experimentally.~\cite{nitta,grundler}
Consequently, much interest has been attracted to the realization
of spin polarized transistors and other devices based on using
electric gate to control the spin dependent transport.~\cite{gt}

Besides using a static gate to control the SOI strength and so control
the stationary spin transport, new physical phenomena can be observed in
time dependent spin transport under the influence of a fast varying gate
voltage. Along this line, in this article we will consider a mechanism of
AC spin current generation using time dependent gate. This mechanism employs
a simple fact that the time variation of Rashba SOI creates a force which
acts on opposite spin electrons in opposite directions. Inversely, when a
gate is coupled to a nearby electron gas, the spin current in this electron
gas also induces a variation of the gate voltage, and hence affects the
electric current in the gate circuit. We will use a simple model to clarify
the principle of such a new detection mechanism without any optical or
magnetic mediator. The systems to be studied will be 1D electron gas in a
semiconductor quantum wire (QWR) and 2D electron gas in a semiconductor
quantum well (QW).

We consider a model in which the Rashba SOI is described by the
time dependent Hamiltonian
$H_{so}(t)=\hbar\alpha(t)(\vec{k}\times\hat{\nu})\cdot\vec{s}$,
where $\vec{k}$ is the wave vector of an electron, $\hbar\vec{s}$
is the spin operator, and $\hat{\nu}$ is the unit vector. For a
QWR $\hat{\nu}$ is perpendicular to the wire axis, and for a QW
perpendicular to the interfaces. The time dependence of the
coupling parameter $\alpha (t)$ is caused by a time dependent
gate.~\cite{dress} To explain clearly the physical mechanisms
leading to the spin current generation, we will first consider the
1D electron gas in a QWR, and assume $\alpha(t)$ to be a constant
$\alpha$ for $t$$<$0, and $\alpha(t)$=0 for $t$$>$0. For the 1D
system we choose the $x$ direction as the QWR axis and $y$ axis
parallel to $\hat{\nu}$, to write the SOI coupling in the form
$H_{so}(t)$=$\hbar\alpha(t)k_xs_z$. For $t$$<$0 the spin
degeneracy of conduction electrons is lifted by SOI, producing a
splitting $\Delta$=$\hbar\alpha k_x$ between $s_{z}$=$1/2$ and
$s_{z}$=$-1/2$ bands, as shown in Fig.~1 by solid curves together
with the Fermi energy $E_F$. The spin current in this state is
zero, as it should be under thermal equilibrium.

Indeed, the spin current is defined as
$I_s(t)$=$I_{\uparrow}(t)-I_{\downarrow}(t)$, where
$I_{\uparrow}(t)$ [or $I_{\downarrow}(t)$] is the partial current
associated with the spin projections $s_z$=$1/2$ (or
$s_z$=$-1/2$). Hence,
\begin{equation}\label{Is}
I_s(t) = \frac{\hbar}{2L} \sum_{E(k_x) < E_F}
[v_{\uparrow}(k_x) - v_{\downarrow}(k_x) ] \, ,
\end{equation}
where $L$ is the length of the QWR. Taking the momentum derivative of the
Hamiltonian, we obtain the velocity as
\begin{equation}\label{v}
v_{\uparrow,\downarrow}(k_x) =
\hbar k_x / m^{\ast} \pm \alpha (t)/2 \, .
\end{equation}
The spin current is then readily obtained as
\begin{equation}\label{Is2}
I_s(t) =  (\hbar n/4 m^{\ast}) (\hbar k_{\uparrow} -
\hbar k_{\downarrow}) + \hbar\alpha(t) n/4 \, ,
\end{equation}
where $n$ is the 1D electron density, and $k_{\uparrow}$ (or
$k_{\downarrow}$) is the average momentum in the $\uparrow$-spin (or
$\downarrow$-spin) band.

For a parabolic band $\hbar k_{\uparrow}$=$-m^{\ast}\alpha/2$ and
$\hbar k_{\downarrow}$=$m^{\ast}\alpha/2$. Although $\hbar
k_{\uparrow}-\hbar k_{\downarrow}$ gives a finite contribution to
$I_s(t)$ in (\ref{Is2}), for $t$$<$0 where $\alpha(t)$=$\alpha$,
this contribution is compensated by the contribution $\hbar\alpha
n/4$ due to the SOI. Hence, the total spin current $I_s(t)$=0 for
$t$$<$0. However, when the SOI is switched off at $t$=0,
$\alpha(t)$=0 and so the spin current is finite, because the
average electron momenta retain the same as they were at $t$$<$0.
As time goes on, the electron momenta relax with a relaxation time
$\tau$. Therefore, $I_s(t)$=$-(\hbar\alpha n/4)\exp (-t/\tau)$ for
$t>0$.
\begin{figure}[bp]
\includegraphics[width=5cm]{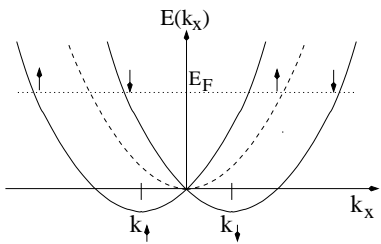}
\caption{The dashed curve is the electron energy band without SOI.
The SOI splits the energy band into the $\uparrow$-spin and the
$\downarrow$-spin band, as shown by the solid curves, with
corresponding average wave vector $k_{\uparrow}$ and
$k_{\downarrow}$.} \label{fig1}
\end{figure}

It is instructive to make a Fourier transform of $I_s(t)$ to obtain a
Drude-like expression
\begin{equation}\label{drude}
I_s(\Omega) = [ \frac{\tau\hbar n}{2 m^{\ast} (i\Omega\tau -1)}] \,
[ \frac{m^{\ast}}{2} i \Omega \, \alpha(\Omega) ] \, .
\end{equation}
Since the units of our spin current is $\hbar/2$, the above expression is a
complete analogy to the electric conductivity. Instead of an electric
driving force $eE$, here we have an equivalent driving force
$(m^{\ast}/2)[d\alpha(t)/dt]$, the Fourier component of which is
$(m^{\ast}/2)\,i\Omega\,\alpha(\Omega)$.
Under this driving force we have  the classic equation of motion
\begin{equation}\label{force}
m^{\ast} \frac{d\,v_{\uparrow,\downarrow}}{dt} =
\pm \frac{m^{\ast}}{2}\frac{d\alpha(t)}{dt}
\end{equation}
This force acts in opposite directions on electrons with opposite spin
projections. When such a force creates a spin current, it does not induce an
electric current.

The above conclusion of spin current generation can be demonstrated with a
rigorous linear response analysis, which will be performed on a 2D electron
gas (2DEG). The simple Drude expression (\ref{drude}) will then appear as a
general result. Let the 2DEG be in the $xy$ plane with the unit vector
$\hat{\nu}$ along the $z$ axis, which is the spin quantization axis. We will
use the equation of motion for the spin density operator to generalize the
1D expressions (\ref{Is}-\ref{v}) for the spin current. For a homogeneous
system the spin current density operators can be expressed in terms of the
electron creation operator $c^{\dag}_{\vec{k},\gamma}$ and destruction
operator $c_{\vec{k},\gamma}$, where $\gamma$ labels the spin projection onto
the z-axis. This current is then derived as
\begin{equation}\label{J+J}
\mathcal{J}_j^i = J_j^i + J^i_{j,_{\rm SOI}} \, ,
\end{equation}
where the superscript $i$=$x,y,z$ specifies the direction of spin polarization,
and the subscript $j$=$x,y$ refers to the direction of the spin current flow.
The partial current
\begin{equation}\label{kin}
J_j^i = \sum_{\vec k} \sum_{\gamma\beta} \frac{\hbar^{2}k_j}{m^*}\,
c^{\dag}_{\vec{k},\gamma}\, s_{\gamma\beta}^i\, c_{\vec{k},\beta}
\end{equation}
is the ordinary kinematic term, and
\begin{equation}\label{jsoi}
J_{j,_{\rm SOI}}^i = \varepsilon^{ijz} \, \hbar\alpha \,n/4
\end{equation}
is the contribution of SOI.~\cite{sch} Here  $\varepsilon^{ijz}$
denotes the Levy-Civita symbol. The SOI induced current resembles
the diamagnetic current of electrons under the action of an
external electromagnetic vector potential.

We note that the SOI Hamiltonian can be conveniently written in terms of the
kinematic current as
\begin{equation}\label{int}
H_{so}(t) = [m^*\alpha(t)/\hbar] \, (J^x_y - J^y_x) \, .
\end{equation}
When an AC bias with frequency $\Omega$ is applied to the front or
the back gate of a 2DEG,~\cite{nitta,grundler} the Rashba coupling
constant contains two terms
$\alpha(t)$=$\alpha_0$+$\delta\alpha(t)$, where $\alpha_0$ is
constant in time and $\delta\alpha(t)$=$\delta\alpha\,e^{i\Omega
t}$. We assume that the only effect of the AC bias is to add a
time dependent component to the SOI coupling constant, although in
practice it is not simple to avoid the bias effect on the electron
density.~\cite{grundler} The SOI Hamiltonian is separated
correspondingly into two parts
$H_{so}(t)$=$H_{so}^0$+$H_{so}^{\prime}(t)$. The time independent
part $H_{so}^0$ does not produce a net spin current in the
thermodynamically equilibrium state. However, as pointed out in
the above analysis on the 1DEG system, the time dependent
$H_{so}(t)$ can give rise to a spin current.

We will incorporate $H_{so}^0$ into our unperturbed Hamiltonian and treat
$H_{so}^{\prime}(t)$ within the linear response regime.
The so-generated AC spin
current $\langle\mathcal{J}^i_j(t)\rangle$ has the form
\begin{eqnarray}\label{Jij}
\langle\mathcal{J}^i_j(t)\rangle &=&
\frac{i}{\hbar} \int\limits_{-\infty}^t dt^{\prime}
\langle \left[ H_{so}^{\prime}(t^{\prime}),J_j^i(t) \right] \rangle \nonumber \\
&& + \, \varepsilon^{ijz} \, \hbar \, \delta\alpha(t)\, n/4 \, .
\end{eqnarray}
In the above equation the first term can be written in the form
$\delta\alpha(t)\mathcal{R}_j^i(\Omega)$. For zero temperature and with
$\Omega$$>$0, the response function $\mathcal{R}_j^i(\Omega)$ can be represented
as the Fourier transform of the correlator
\begin{eqnarray}\label{response}
\mathcal{R}^i_j(t) &=& -i\frac{\hbar^2}{m^*}
\sum_{\vec{k}^{\prime}\alpha^{\prime}\beta^{\prime}}
k_j^{\prime}\,s^i_{\alpha^{\prime}\beta^{\prime}}
\sum_{\vec{k}\alpha\beta} \vec{h}_{\vec{k}} \cdot \vec{s}_{\alpha\beta} \\
&&\times \,\, \overline{ \langle T \{
c^{\dag}_{\vec{k}^{\prime}\alpha^{\prime}}(t) \, c_{\vec{k}\beta} \} \rangle \,
\langle T \{ c_{\vec{k}^{\prime}\beta^{\prime}}(t) \,
c^{\dag}_{\vec{k}\alpha} \} \rangle } \, , \nonumber
\end{eqnarray}
where $\vec{h}_{\vec{k}}$=$\vec{k}$$\times$$\hat{\nu}$. In the above equation,
the bar over the product of two one-particle Green functions means an ensemble
average over impurity positions.

We will use the standard perturbation theory~\cite{alt} to
calculate this ensemble average, which is valid when the elastic
scattering time $\tau$ due to impurities is sufficiently long such
that $E_F\tau$$\gg$$\hbar$. We will assume that the electron Fermi
energy $E_F$ is much larger than both $\hbar\Omega$ and
$\hbar\alpha_0 h_{\vec{k}}$. To the first order approximation, we
neglect the weak localization corrections to the correlator
(\ref{response}), since these corrections simply renormalize the
spin diffusion constant.~\cite{mc} Consequently, the configuration
average of the pair product of Green functions is expressed in the
so-called ladder series.~\cite{alt} We found that since
$\vec{h}_{\vec{k}}= -\vec{h}_{-\vec{k}}$ many of such ladder
diagrams vanish after angular  integration in (\ref{response}),
similar to suppression of ladders in the electric current driven
by the vector-potential.~\cite{alt} At the same time, some of
nondiagonal on spin indices diagrams do not turn to 0 after the
angular integration. Employing the analysis of similar diagrams
done in it can be shown that they cancel each other.~\cite{mc}
Hence,  the configuration average in (\ref{response}) decouples
into a product of average Green functions and
Eq.~(\ref{response}) becomes
\begin{eqnarray}\label{corr}
\mathcal{R}_j^i(\Omega) &=& -i \frac{\hbar^2}{m^*} \sum\limits_{l,n}
\varepsilon^{lnz} \sum\limits_{\vec{k}} k_jk_n \\
&&\times \int \frac{d\omega}{2\pi} \, {\mathop{\rm Tr}\nolimits}
\left[ s^l G(\vec{k},\omega) \, s^i G(\vec{k},\omega+\Omega) \right] \, ,
\nonumber
\end{eqnarray}
where $G(\vec{k},\omega)$ is the average Green's function which contains fully
the effect of $H_{so}^0$. This function is represented by the 2$\times$2 matrix
\begin{eqnarray}\label{G}
G(\vec{k},\omega) = \left[ \omega - E_{\vec{k}}/\hbar -
\alpha_0\vec{h}_{\vec{k}} \cdot \vec{s} + i \, \Gamma {\rm sgn}(\omega)
\right]^{-1} \, ,
\end{eqnarray}
where $\Gamma$=1/2$\tau$, and $E_{\vec{k}}$  is defined with respect to $E_F$.
Substituting (\ref{G}) into (\ref{corr}), and then
into (\ref{Jij}), we obtain the spin current
\begin{equation}\label{final2}
\langle \mathcal{J}^i_j(\Omega) \rangle = \varepsilon^{ijz} \,
\frac{\hbar}{4} \, \delta\alpha \,n \frac{\Omega}{\Omega+2i\Gamma} \, .
\end{equation}
It is important to point out that the spin density under the gate area is zero.
This is the reason why even in a 2DEG the D'yakonov-Perel spin
relaxation~\cite{dp} does not appear in (\ref{final2}) for the generated spin
current, although this spin current is determined by the response function
(\ref{corr}) which involves spin degrees of freedom. Hence, in the homogeneous
system with zero spin density, only electron momentum relaxation occurs in the
process of spin current generation by a time dependent gate.

Unlike the spin current (\ref{drude}) in a 1D system, in a 2DEG the current
given by (\ref{final2}) has no specific direction. To clarify the spatial
distribution of the spin flux induced by an AC gate, let us take the chiral
component $\mathcal{J}_{\rm chir}(t)$ of the spin current
\begin{equation}\label{chir}
\mathcal{J}_{\rm chir}(t) = [\, \langle\mathcal{J}^x_y(t)\rangle -
\langle\mathcal{J}^y_x(t)\rangle \, ]/2 \, .
\end{equation}
It is easily to see that this chiral projection has the same form
as the expression (\ref{drude}) for a 1D system, if $n$
represents the electron density of the 2DEG. In Fig.~2 we
illustrate the spin current distribution for a circular gate
which is marked as the gray area. The spin polarization at any
point under the gate has two components parallel to the 2DEG. For
any direction specified by the unit vector $\vec{N}$, the two
spin polarized flux with polarization directions parallel and
antiparallel to $\vec{N}$ will oscillate out of phase by the
amount of $\pi$  along the direction perpendicular to $\vec{N}$.
Such out of phase oscillation is schematically plotted in Fig.~2.
The amplitude of the spin density flow in each of the opposite
directions, as marked by the dashed-line arrows, is just
$\mathcal{J}_{\rm chir}(t)$. In the 2DEG outside the gate area,
the spin current can be supported only by spin diffusion.
Therefore the chiral AC spin polarization is accumulated in the
vicinity of the circumference of the gate, and from where
diffuses away from the gate area. It can also diffuse under the
gate. For small gates such back diffusion can diminish the
efficiency of the spin generation. On the other hand, for large
gates with the size larger than the spin diffusion length the
diffusion counterflow does not reduce much the total spin current.
\begin{figure}[tbp]
\includegraphics[width=3.7 cm]{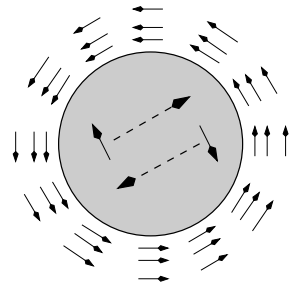}
\caption{Distribution of spin currents induced by a time dependent circular
gate which is marked as the gray region. Under the gate, electrons with
opposite spins (solid arrows) move in opposite directions indicated by the
dashed-line arrows.
Arrows outside the gate area show the accumulated spin polarization during a
half period of AC gate voltage oscillation.}
\label{fig2}
\end{figure}

The so-generated current amplitude can be easily estimated. With
$\delta\alpha$=3$\times$10$^6$ cm/s,~\cite{grundler} for
$\Omega$=$2\pi$$\times$10$^9$ s$^{-1}$, $n$=10$^{12}$ cm$^{-2}$,
and $\tau$=1 ps, from (\ref{final2}) we derive
$(2e/\hbar)\langle\mathcal{J}^i_j(\Omega)\rangle$$\simeq$10$^{-3}$
Amp/cm. This AC spin current can be detected by various methods.
For example, if holes can tunnel into the neighborhood of the gate
edge, their recombination with spin polarized electrons will
produce the emission of circular polarized light~\cite{ohno}.

However, here we will discuss a new method of direct electric detection of the
DC or the AC spin current. This method is based on a simple fact that the
Rashba SOI couples the spin current to the gate voltage. We have shown in our
above analysis that due to this coupling, spin current can be induced by a time
dependent gate voltage. In this case the voltage variation plays the role of a
source which drives electrons out of thermodynamic equilibrium, and the spin
current is the linear  response to this perturbation. The reverse process is to
create a spin current in a 2DEG by some source, and so inducing a voltage shift
in a nearby gate. This is also possible to realize. We thus consider a model
where the SOI constant $\alpha(U)$ is a function of the gate voltage
$U(t)$=$U_0$+$V(t)$. $U_0$ is the static equilibrium value in the absence of a
spin current, while $V(t)$ is a dynamic variable. The mean value
$\langle V \rangle$  of $V(t)$ has to be calculated as a linear response to the
perturbation associated with the presence of the spin polarization flow. The
explicit form of this  perturbation can be obtained by averaging the
Hamiltonian of the system over an electronic state with the given time
dependent spin current.

Let $\langle\cdots\rangle_J$ be such type of average. To the lowest order with
respect to SOI, the coupling of the gate voltage to the spin current is thus
determined by the average of the Rashba interaction in (\ref{int}) with
$\alpha$=$\alpha(U)$. The coupling between the gate voltage $U(t)$ and the
spin current $\mathcal{J}_j^i$ is via the kinetic current ${J}^i_j$. To derive
the coupling Hamiltonian $H_{int}$, we use (\ref{J+J}) to express ${J}^i_j$ in
terms of $\mathcal{J}_j^i$, and expand
$\alpha(U)$=$\alpha(U_0)$+$\alpha^{\prime}V(t)$ for small $V(t)$.
The coupling
Hamiltonian is then derived from (\ref{int}) as
\begin{equation}\label{jvint}
H_{int} = \frac{m^*\alpha^{\prime}}{\hbar} V \,
[\, \langle\mathcal{J}^x_y\rangle_J - \langle\mathcal{J}^y_j\rangle_J \,] \, .
\end{equation}
The charging of the gate $Q$=$CV$ is related to the gate capacitance. Hence,
(\ref{jvint}) can be expressed in the convenient form $H_{int}$=$Q\mathcal{E}$,
where
\begin{equation}\label{emf}
\mathcal{E} = \frac{m^*\alpha^{\prime}}{\hbar C}
[\, \langle\mathcal{J}^x_y\rangle_J - \langle\mathcal{J}^y_j\rangle_J \,]
\end{equation}
is the effective electromotive force.
\begin{figure}[tbp]
\includegraphics[width=.33 \textwidth,angle=0]{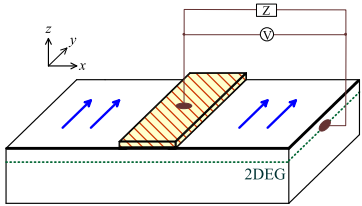}
\caption{Schematic illustration of spin current detection. AC spin
current flows from the right to the left under the gate with spin
polarized as shown by arrows. $V$ denotes the voltmeter and $Z$ is
the outer circuit impedance.} \label{fig3}
\end{figure}

To illustrate our proposed method of direct electric detection,
let us consider a  circuit connected to the gate.  The principal
scheme of the spin current detection is shown in Fig.~\ref{fig3}.
In it, an additional back gate can be utilized to tune the
electron density (not shown).
 The circuit is characterized by a frequency dependent impedance
$Z(\Omega)$. The voltage induced on the gate by the electromotive
force (\ref{emf}) is then easily obtained as
\begin{equation}\label{v2}
\langle V \rangle = \mathcal{E}
\frac{i\Omega CZ(\Omega)}{1+ i\Omega CZ(\Omega)} \, .
\end{equation}
When the spin current frequency is in resonance with the circuit
eigenmode, the gate voltage becomes very large. In the limit of
high impedance (open circuit), $\langle V \rangle$=$\mathcal{E}$.
Using the spin current
$(2e/\hbar)\langle\mathcal{J}^i_j(\Omega)\rangle$$\simeq$10$^{-3}$
Amp/cm derived above, and the fact that
$\langle\mathcal{J}^i_j\rangle_J=A\,\langle\mathcal{J}^i_j(\Omega)\rangle$,
where $A$ is the area under the gate, let us estimate the
electromotive force induced in a probe gate by this spin current
generated by a nearby source gate. For the reasonable parameter
values $\alpha^{\prime}$=3$\times$10$^7$ cm/Vs,~\cite{grundler}
$m^*$=0.03 $m_{e}$, and $C$=$\kappa\epsilon_0\,A/l$ with
$\kappa$=10 and $l$=10$^{-5}$ cm, from (\ref{final2}) and
(\ref{v2}) we obtain $\mathcal{E}$$\simeq$10$^{-5}$ Volts.

The generated AC spin current can be rectified with various methods. For
example, one can use a shutter gate which is $\pi/2$ phase shifted with respect
to the generation gate. The shutter gate can be placed in the neighborhood of
the generation gate or between two such gates. The evaluation of the rectifying
efficiency  of such a setup requires a thorough analysis of spin relaxation and
diffusion processes caused by the spin accumulation during the shutter cycle.

We would like to add one relevant information which we became
aware of after we have completed this paper. The preprint of
Governale {\it et al.} on the quantum spin pumping in a 1D wire is
also based on the idea of creating spin current via time dependent
gate.~\cite{governale} However, our results involving dissipative
transport in 2DEG and 1DEG can not be compared directly with those
in Ref.~\onlinecite{governale}.

This work was supported by the National Science Council of Taiwan
under grant Nos. 91-2119-M-007-004 (NCTS), 91-2112-M-009-044
(CSC), the Swedish Royal Academy of Science, and the Russian
Academy of Sciences and the RFBR grant No. 03-02-17452. AGM
acknowledges the hospitality of NCTS in Hsinchu where this work
was initiated.


\begin{thebibliography}{99}
\bibitem{dress}
We ignored the bulk Dresselhaus contribution to
SOI, which in narrow gap heterostructures is believed to
be negligible in comparison with the Rashba effect. See
discussion in [4].
\end{thebibliography}
\end{document}